\documentclass[print]{pasj01}
\begin{document} 
\Received{}%{yyyy/mm/dd}
\Accepted{}%{yyyy/mm/dd}
%\Published{yyyy/mm/dd}

\title{Failed Supernova Remnants}

%%% begin:list of authors
% Do NOT capitalize all letters in "textsc".
\author{Daichi \textsc{Tsuna}\altaffilmark{1,2}
\email{tsuna@resceu.s.u-tokyo.ac.jp}}
\altaffiltext{1}{Research Center for the Early Universe (RESCEU), School of Science, The University of Tokyo, 7-3-1 Hongo, Bunkyo-ku, Tokyo 113-0033, Japan}
\altaffiltext{2}{Department of Physics, School of Science, The University of Tokyo, Tokyo, Japan}

%%% end:list of authors

%% `\KeyWords{}' always has to be placed before `\maketitle'.
\KeyWords{stars: black holes --- ISM: supernova remnants --- supernovae: general} %Do NOT move this preamble from here!

\maketitle

\begin{abstract}
In a failed supernova, partial ejection of the progenitor's outer envelope can occur due to weakening of the core's gravity by neutrino emission in the protoneutron star phase. We consider emission when this ejecta sweeps up the circumstellar material, analogous to supernova remnants (SNRs). We focus on failed explosions of blue supergiants, and find that the emission can be bright in soft X-rays. Due to its soft emission, we find that sources in the Large Magellanic Cloud (LMC) are more promising to detect than those in the Galactic disk. These remnants are characteristic in smallness ($\lesssim 10$ pc) and slowness (100s of ${\rm km\ s^{-1}}$) compared to typical SNRs. %i.e. they will look like young SNRs in terms of size but with a narrow (100s of ${\rm km\ s^{-1}}$) width in their line emission. 
Although the expected number of detectable sources is small (up to a few by eROSITA 4-year all-sky survey), prospects are better for deeper surveys targeting the LMC. Detection of these ``failed SNRs" will realize observational studies of mass ejection upon black hole formation.
\end{abstract}

\section{Introduction}
Stellar mass black holes (BHs) are nowadays routinely found as X-ray \citep{Remillard06,Corral16} and gravitational-wave \citep{Abbott19,Abbott20} sources. A major pathway to form them is believed to be the gravitational collapse of massive stars. This is proposed to explain the absence of supernova (SN) progenitors in certain mass ranges (e.g. \cite{Kochanek08,Smartt09}), as well as the mass distribution of observed BHs \citep{Kochanek14,Raithel18}. Theoretical works also find that stars having compact cores likely fail to revive the shock formed upon core bounce (e.g. \cite{OConnor11,Sukhbold16,Ertl16}), resulting in ``failed supernovae" that leave behind BHs. 

Even if the bounce shock cannot propagate, some mass can still be ejected from the collapsing star. At the protoneutron star phase preceding BH formation, the core decreases its gravitational mass due to neutrino emission, by up to a few 10\% in a few seconds (e.g., \cite{OConnor13}). The sudden loss of gravity creates a sound pulse, which steepens into a shock and can eventually shed the outer envelope once it propagates to the star's surface \citep{Nadyozhin80,Lovegrove13,Fernandez18,Coughlin18a,Coughlin18b}.

Table \ref{table:Parameters} shows the parameters of the ejecta obtained from recent hydrodynamical simulations of mass ejection, for different types of progenitors (\cite{Fernandez18,Tsuna20}; see also \cite{Ivanov21}). The dependence of the ejecta parameters on progenitor types can be understood from the compactness of their envelopes. Red supergiants (RSGs) have loose envelopes, which realize ejection of a good fraction of its envelope but with a slow velocity. Wolf-Rayet stars (WRs) have compact envelopes with large binding energy, which makes the ejecta very light but fast. The parameters of the ejecta from blue supergiants (BSGs) lie in the middle of the two.

When the ejecta collide with the circumstellar medium (CSM), shock-heated gas generates emission analogous to supernova remnants (SNRs). The focus of this study is to understand the emission of these ``failed supernova remnants", and estimate the detectability of the sources in the Galaxy and the Large Magellanic Cloud (LMC). Observation of failed SNRs will be important to uncover the properties of mass ejection upon BH formation, for which observational clues are scarce at present.

In this work we mainly consider failed SNRs from BSGs, and find that they may be detectable as soft X-ray sources. We find that the all-sky survey by eROSITA \citep{Merloni12} may detect up to a few sources, likely in the LMC.

This letter is constructed as follows. In Section \ref{sec:methods} we estimate the event rate of these sources, and present our methods for obtaining the X-ray light curve. We show our results in Section \ref{sec:results}. In Section \ref{sec:discussion} we apply our results to estimate the detectability, and briefly comment on failed SNRs from progenitors other than BSGs.

\begin{table}
\centering
\begin{tabular}{c|ccc}
Model & $M_{\rm ej}$ [$M_\odot$] & $E_{\rm ej}$ [erg] & $v_{\rm ej}$ [km s$^{-1}$] \\ \hline
RSG & $4$ & $2\times 10^{47}$ & 70 \\
BSG-1 & $0.1$ & $6\times 10^{47}$ & 800\\
BSG-2 & $0.05$ & $2\times 10^{47}$ & 630 \\
WR & $5\times 10^{-4}$& $3\times 10^{46}$ & 2000 
\end{tabular}
\caption{Typical parameters of ejected material from a failed supernova, for each type of star. Columns are: type of massive star, ejecta mass, ejecta kinetic energy, and the velocity ($\sqrt{2E_{\rm ej}/M_{\rm ej}}$). The parameters of the ejecta are taken from \citet{Fernandez18}, except for BSG-1 that is from \citet{Tsuna20}.}
\label{table:Parameters}
\end{table}

\section{Failed SNRs by Blue Supergiants}
\label{sec:methods}
\subsection{Event rate}
A major factor that sets the detectability is the event rate of failed SNe from BSGs in our Galaxy and LMC. These are uncertain, but can be roughly estimated as follows. 

The (successful) explosions of BSGs are often tied to Type II-pec SNe, the representative being SN 1987A. This class accounts for around $2\%$ of core-collapse SNe \citep{Smartt09,Kleiser12,Pastorello12}. Assuming this class is the one and only class of BSG explosions and using the Galactic core-collapse SN rate $\sim 3\times 10^{-2}\ {\rm yr^{-1}}$ \citep{Adams13}, the Galactic rate of BSG explosions is $\sim 6\times 10^{-4}\ {\rm yr^{-1}}$. The fraction of failed SNe is estimated for core-collapse of RSGs \citep{Adams17b}, with a median value $\approx 0.14$. However, this fraction is unconstrained for BSGs, and can be much larger. There are several studies that claim BH formation from a BSG could explain (at least a fraction of) optical or radio transients of unknown origin \citep{Kashiyama15,Kashiyama18,Tsuna20}, which all occur at a rate of a few \% of core-collapse SNe. The fact that SN 1987A was a successful SN may imply that the number of failures is at most comparable to the number of successes.  We thus estimate the Galactic rate $\mathcal{R_{\rm MW}}$ of failed BSG explosions to be in the range $\mathcal{R_{\rm MW}}\sim 10^{-4}$--$10^{-3}\ {\rm yr^{-1}}$. For the LMC, home to SN 1987A, the present-day star formation rate is about $0.4\ M_\odot\ {\rm yr^{-1}}$ \citep{Harris09}, which is about 1/5--1/2 of our Galaxy (e.g. \cite{Robitaille10,Davies11,Licquia15}). Thus the corresponding event rate in the LMC, $\mathcal{R}_{\rm LMC}$, is likely to be lower by a similar factor. In summary, we adopt $\mathcal{R}_{\rm MW}=3\times 10^{-4}{\rm yr^{-1}}, \mathcal{R}_{\rm LMC}=10^{-4}{\rm yr^{-1}}$ as representative values, but we should keep in mind the uncertainties in these rates spanning an order of magnitude.

Overall, the timescale of interest is $\sim 10^4$ years; are SNRs from failed BSG explosions detectable at this age?

\subsection{Light Curve Modelling}
For SNRs of core-collapse origin, just outside the star lies the stellar wind emitted prior to core-collapse. Outside this depends on the evolution history of the progenitor (e.g., \cite{Dwarkadas07}). Here we consider a simple profile of three phases, a BSG wind, shell created by mass lost mostly during the preceding RSG phase, and a bubble created by the main-sequence wind. We assume the shell to be homogeneous with total mass $M_{\rm sh}=10M_\odot$, and the outermost bubble to have number density $0.01\ {\rm cm^{-3}}$ and temperature $10^6$ K \citep{Castor75}. The choice of $M_{\rm sh}$ is inspired by the mass lost for the BSG progenitor adopted in \citet{Fernandez18}. The outer extent of the RSG shell is set by the equilibrium between the RSG wind's ram pressure and the pressure in the bubble $p_{\rm bubble}$
\begin{eqnarray}
    r_{\rm 1}&\approx& \left(\frac{\dot{M}_{\rm RSG}v_{w, \rm RSG}}{4\pi p_{\rm bubble}}\right)^{1/2} \nonumber \\
    &\sim& 7\ {\rm pc} \left(\frac{\dot{M}_{\rm RSG}v_{w,\rm RSG}}{10^{-3}M_\odot{\rm yr^{-1}} {\rm km\ s^{-1}}}\right)^{1/2}\left(\frac{p_{\rm bubble}}{10^{-12}\ {\rm erg\ cm^{-3}}}\right)^{-1/2}.
\end{eqnarray}
%the ejecta are surrounded by the progenitor's wind, up to some radius where it transitions to an interstellar medium (ISM) of roughly constant number density $n_0$. 
%We adopt a simplified CSM with number density profile (e.g., \cite{Kashiyama18})
%\begin{eqnarray}
%n_{\rm CSM} (r)&=& \left\{ \begin{array}{ll}
%\frac{\dot{M}}{4\pi r^2 v_w m_p} &(r<r_{\rm eq}), \\
%\dot{M}/(4\pi r^2 v_w m_p) & (r<r_{\rm eq}), \\
%n_0 &(r>r_{\rm eq}),
%\end{array}\right.
%\end{eqnarray}
%where $\dot{M}$ and $v_w$ are respectively the mass loss rate and the wind velocity. 
We adopt $\dot{M}_{\rm RSG}=3\times 10^{-5}M_\odot\ {\rm yr}^{-1}, v_{w,\rm RSG}=25\ {\rm km\ s^{-1}}$. The inner extent $r_2$ is set to where the BSG wind's ram pressure is equivalent to $p_{\rm bubble}$
\begin{eqnarray}
r_{\rm 2} \sim 2\ {\rm pc} \left(\frac{\dot{M}v_w}{10^{-4}M_\odot\ {\rm yr^{-1}}{\rm km\ s^{-1}}}\right)^{1/2} \left(\frac{p_{\rm bubble}}{10^{-12}\ {\rm erg\ cm^{-3}}}\right)^{-1/2}.
\label{eq:r_eq}
\end{eqnarray}
where $\dot{M}, v_w$ are the mass loss rate and wind velocity at the BSG phase. The number density in the shell is given as $3M_{\rm sh}/4\pi m_p(r_1^3-r_2^3)$,
%\sim 1\ {\rm cm^{-3}}(M_{\rm sh}/10M_\odot)[{(r_1^3-r_2^3)}/{100\ {\rm pc^3}}]^{-1}$
where $m_p$ is the proton mass.

For BSGs $v_w$ is of order $100\ {\rm km\ s^{-1}}$, but $\dot{M}$ can have a large variation. 
For the BSG progenitor used in \citet{Fernandez18} and \citet{Tsuna20}, the effective temperature and radius are $15000$ K and $100R_\odot$ respectively. A recent model of line-driven stellar wind \citep{Krticka21} predicts BSGs with these parameters have winds of $v_w\approx100\ {\rm km\ s^{-1}}$ and $\dot{M}\approx 1.5\times 10^{-6}M_\odot\ {\rm yr}^{-1}$.
For simplicity we fix $v_w$ as $100\ {\rm km\ s^{-1}}$, and vary $\dot{M}$ as four values around the above prediction: $\{0.3,1,3,6\}\times 10^{-6}M_\odot\ {\rm yr^{-1}}$. %We adopt the standard values for the ISM parameters, $n_0=1\ {\rm cm^{-3}}$ and $c_s=10\ {\rm km\ s^{-1}}$.

We consider BSG ejecta with parameters of BSG-1, BSG-2 in Table \ref{table:Parameters}. Initially the ejecta are heavier than the swept-up wind, and they are in a coasting phase. This continues until the ejecta sweep up CSM equal to its own mass, from which the ejecta decelerate (Sedov phase). During these adiabatic phases, the radius and velocity of the forward shock $r_{\rm sh}, v_{\rm sh}$ evolve by the following set of equations
\begin{eqnarray}
v_{\rm sh}(t+\Delta  t) &=& v_{\rm sh}(t) \left[\frac{M_{\rm ej}+\delta M(t)}{M_{\rm ej}+\delta M(t+\Delta t)}\right]^{1/2} \label{eq:vofdeltaM} \\
\delta M(t+\Delta t) &=& \delta M(t) + 4\pi r_{\rm sh}^2 n_{\rm CSM}m_p {\rm max}(v_{\rm sh}-v_w, 0)\Delta t \label{eq:deltaMoft} \\
r_{\rm sh}(t+\Delta t) &=& r_{\rm sh}(t) + v_{\rm sh} \Delta t, 
\end{eqnarray}
where $\delta M$ is the swept-up mass. When the shock is radiative, equation (\ref{eq:vofdeltaM}) becomes a momentum-conserving one
\begin{eqnarray}
v_{\rm sh}(t+\Delta  t) &=& v_{\rm sh}(t) \left[\frac{M_{\rm ej}+\delta M(t)}{M_{\rm ej}+\delta M(t+\Delta t)}\right].
\end{eqnarray}
The initial condition is set to be $\delta M=0$, $v=v_{\rm ej}$, and $r=100\ R_{\odot}$, the last being the typical radius of BSGs. As particles in the shell move at random directions, $v_w$ is set to zero after the shock enters the shell. When the shock exits the shell, $v_{\rm sh}<100$ km s$^{-1}$, and will thus merge with the bubble. We set $v_{\rm sh}$ to zero from this epoch.

We assume that a strong shock forms between the ejecta and the CSM, and that equipartition between electrons and ions will be achieved. The shock is assumed to be adiabatic with index $5/3$ (compression ratio $4$) throughout, although this can change during the evolution. The temperature of immediate downstream at time $t$ is then obtained from the jump conditions as
\begin{eqnarray}
k_B T_{\rm d}(t)&=&\frac{3}{16}\mu m_p[v_{\rm sh}(t)-v_w]^2\nonumber \\
&\sim& 0.6\ {\rm keV}\left(\frac{\mu}{0.62}\right)\left(\frac{v_{\rm sh}(t)-v_w}{700\ {\rm km\ s^{-1}}}\right)^2    
\end{eqnarray}
where $k_B$ is the Boltzmann constant and $\mu$ is the mean molecular weight, which we set to the solar metallicity value $0.62$. We thus expect emission in soft X-rays, likely dominated by line emission. The cooling timescale for a CSM swept up at time $t$ is
\begin{eqnarray}
t_{\rm cool}(t)&=& \frac{1.5(4n_{\rm CSM})k_BT_{\rm d}(t)}{(4n_{\rm CSM})^2\Lambda} \nonumber \\
&\sim& 10^5\ {\rm yr}\left(\frac{n_{\rm CSM}}{1\ {\rm cm^{-3}}}\right)^{-1}\left(\frac{k_BT_{\rm d}}{0.6\ {\rm keV}}\right) 
%\nonumber \\
%&& \times
\left(\frac{\Lambda(T_{\rm d})}{10^{-22}{\rm cgs}}\right)^{-1}
\end{eqnarray}
where $\Lambda(T)$ is the temperature-dependent cooling function.  We use the tabulated data of the cooling function obtained in \citet{Schure09}, which assumes collisional ionization equilibrium and solar metallicity.%The value of $\Lambda$ at $k_BT_d=0.6$ keV is $\approx 5\times 10^{-23}{\rm erg\ cm^3\ s^{-1}}$, while it increases for $k_BT_d<0.1$ keV.

We calculate the luminosity as follows. At a given time $t$, the cooling timescale defines an adiabatic region in the downstream, i.e. the CSM which crossed the shock at time $t'$, radius $r'$ that still contributes to the luminosity at $t$. The luminosity is defined as the integral within this region
% \begin{eqnarray}
% L_X(t)&\approx& \int 4\pi r^2 dr \left(\frac{r_{\rm sh}}{r}\right)^3 \times \left[4n_{\rm CSM}(r)\cdot\left(\frac{r_{\rm sh}}{r}\right)^{-3}\right]^2 \nonumber \\
% &\times& \left[\eta_X\left(T_{\rm d}\left(\frac{r_{\rm sh}}{r}\right)^{-2}\right)\Lambda\left(T_{\rm d}\left(\frac{r_{\rm sh}}{r}\right)^{-2}\right)\right]
% \end{eqnarray}
\begin{eqnarray}
L_X(t)&\approx& \int 4\pi r'^2 dr' \left(\frac{r_{\rm sh}}{r'}\right)^3 \times \left[4n_{\rm CSM}(r')\cdot\left(\frac{r_{\rm sh}}{r'}\right)^{-3}\right]^2 \nonumber \\
&\times& \left[\eta_X\left(T_{\rm d}(t')\left(\frac{r_{\rm sh}}{r'}\right)^{-2}\right)\Lambda\left(T_{\rm d}(t')\left(\frac{r_{\rm sh}}{r'}\right)^{-2}\right)\right]
\end{eqnarray}
where the factors regarding powers of $(r_{\rm sh}/r')$ are for taking into account adiabatic expansion, and $\eta_X(T)$ is a parameter that determines the fraction of the cooling radiation that goes to the X-ray energy range of interest. X-ray emission is non-negligible for $k_BT\gtrsim 0.1$ keV, while $k_BT$ below this value would result in too weak X-ray emission. The value of $\eta_X$ would also depend on the energy range of the detector; in the next section we will discuss the case for observation by the eROSITA all-sky survey.

\section{Results}
\label{sec:results}
\begin{figure*}
      \centering
      \includegraphics[width=\linewidth]{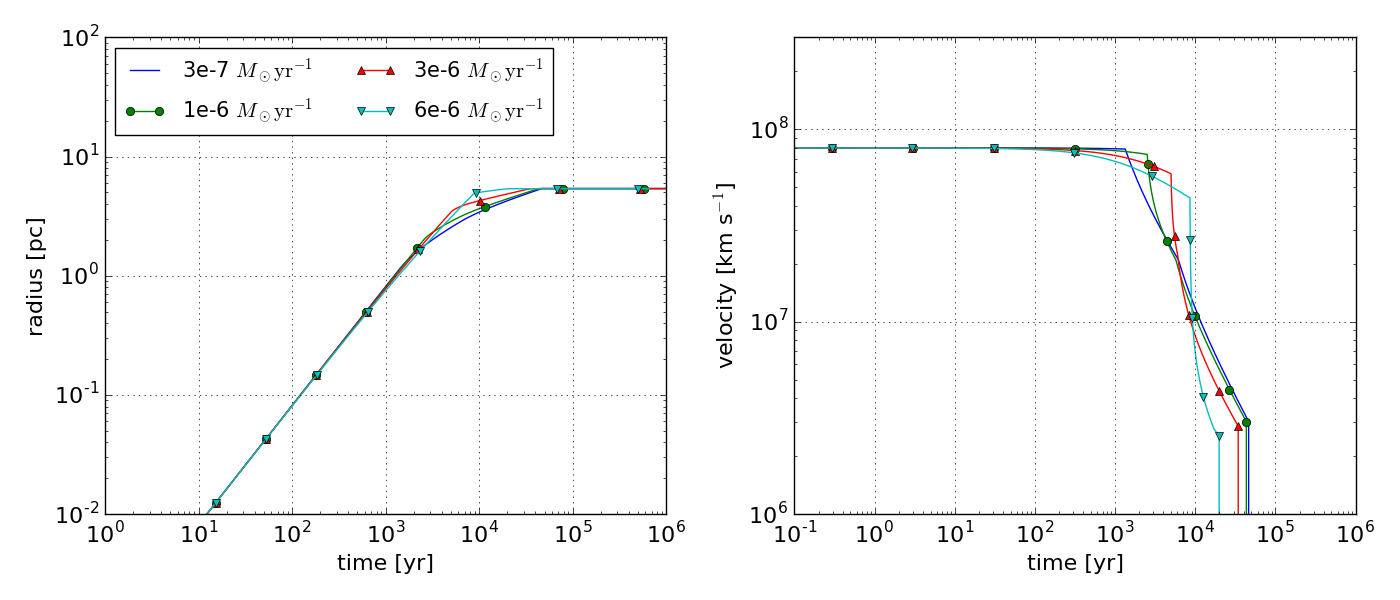}
      \caption{Radius and velocity of the forward shock as a function of time for the BSG-1 model. The different markers indicate different values of $\dot{M}$.}
      \label{fig:dyn_BSG1}
\end{figure*}
We first present the evolution of the shock and the cooling timescale for the BSG-1 model. The evolution is essentially the same for the BSG-2 model, and we do not show them here. We then show the X-ray light curves calculated from our model, for both BSG-1 and BSG-2 models.

The radius and velocity of the shock for the BSG-1 model are in Figure \ref{fig:dyn_BSG1}. The shock evolutions are similar up to $10^3$ years, when the ejecta are in the coasting phase. We see a sudden change in the evolutions when the ejecta reach $r=r_2$, because of the sudden density increase of factor $\sim100$ during the transition to the shell. This occurs later for higher $\dot{M}$, as is clear from equation (\ref{eq:r_eq}).

\begin{figure}
      \centering
      \includegraphics[width=\linewidth]{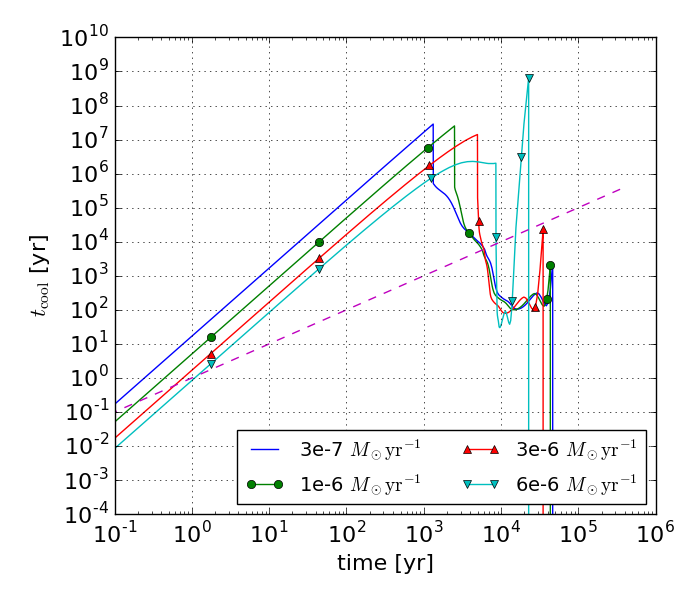}
      \caption{Cooling timescale of the shock downstream as a function of time for the BSG-1 model. The dashed line shows the relation $t_{\rm cool}=t$.}
      \label{fig:cool_BSG1}
\end{figure}

We next discuss the cooling at the immediate downstream. Figure \ref{fig:cool_BSG1} shows the evolution of the cooling time $t_{\rm cool}(t)$. Initially the value of $t_{\rm cool}$ rises, as $t_{\rm cool}\propto n_{\rm CSM}^{-1}\propto r^2$ in the coasting phase. Then $t_{\rm cool}$ suddenly drops when the shock reaches the radius $r=r_{\rm 2}$, due to the sudden rise of the density. The value of $t_{\rm cool}$ further drops as the shock decelerates and $\Lambda$ increases. Then the shock front becomes radiative at around $10^4$ years. From this point the shock velocity has dropped to $\lesssim 100\ {\rm km\ s^{-1}}$, and the immediate downstream will not contribute to the X-ray luminosity. The main contribution is from the plasma heated to X-ray temperatures when the shock was still fast, and have not yet cooled down much by expansion. 

\begin{figure}
    \centering
    \includegraphics[width=\linewidth]{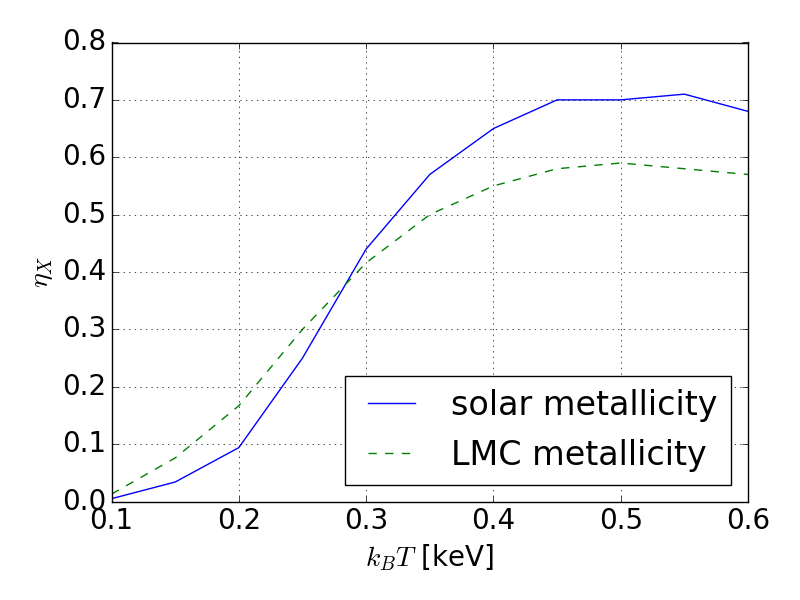}
\caption{The fraction of the bolometric luminosity emitted in the energy range 0.5-2 keV, as a function of temperature $T$. This is obtained from spectra calculated by a numerical code formulated in \citet{Masai94}.}
\label{fig:etaX}
\end{figure}

We estimate the detectability of failed SNRs from BSGs by the eROSITA all-sky survey, which covers an energy range of $0.5$--$2$ keV in the soft X-ray band. To consider this we first calculate X-ray spectra using a spectral synthesis code that takes into account ionization and line emission of elements up to iron \citep{Masai84,Masai94}, and obtain $\eta_X(T)$. $\eta_X(T)$ in the range of interest, for solar and LMC metallicities \citep{Maggi16}, are shown in Figure \ref{fig:etaX}. In this work we use the solar metallicity values for $k_BT>0.1$ keV, and set $\eta_X=0$ otherwise. This may slightly overestimate $L_X$ for the LMC sources, but only by order 10\%\footnote{The adopted $\Lambda$ is a bigger overestimation for LMC sources with lower metallicity. At $t\gtrsim 10^4$yr of interest, the flux is roughly proportional to $\Lambda$.}.

\begin{figure*}
%\begin{tabular}{cc}
% \begin{minipage}{0.5\hsize}
\centering
\includegraphics[width=\linewidth]{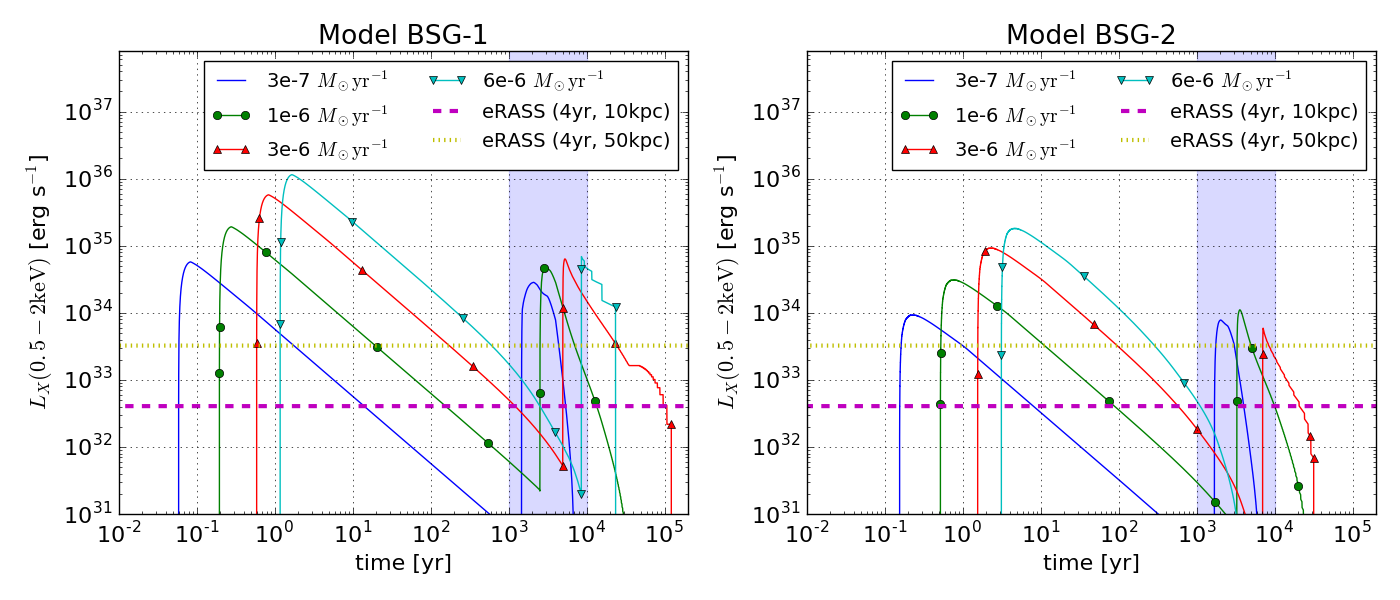} 
%\end{minipage}
% \begin{minipage}{0.5\hsize}
% \centering
%\includegraphics[width=\linewidth]{failedBSG2_lc.png} 
%\end{minipage}
\caption{X-ray light curves of our BSG-1 (left) and BSG-2 (right) models in the energy range $0.5$--$2$ keV. The shaded region shows the inverse of the event rate in the Galaxy. The dashed and dotted lines show the sensitivity of eROSITA all-sky survey (4 years), for a source in the Galaxy and the LMC respectively.}
\label{fig:lc_BSG1}
%\end{tabular}
\end{figure*}

The X-ray light curves calculated using this $\eta_X$ for models BSG-1 and BSG-2 are shown in Figure \ref{fig:lc_BSG1}. We see that the shapes of the curves are similar, with monotonic decay and rebrightening when the forward shock crosses $r=r_2$ and the density increases. The curves for the BSG-1 model generally show higher luminosity than those in the BSG-2 model, since $v_{\rm ej}$ is larger and the downstream plasma can longer maintain X-ray temperatures.

At $t\sim 10^4$ years that we are interested in, the dominant emission thus comes from $r\approx r_2$.
The angular resolution of eROSITA is 30 arcseconds \citep{Merloni12}. This corresponds to a diameter of $1.5$ pc for a hypothetical Galactic source 10 kpc away, while it is $7.5$ pc for sources in the LMC that are 50 kpc away. Since $r_2$ is about a few pc, Galactic sources are likely extended while LMC sources are marginally point sources. The sensitivity of  eROSITA's 4-year all-sky survey in the energy range $0.5-2$ keV is $1.1\ (3.4)\times 10^{-14}\ {\rm erg\ s^{-1}\ cm^{-2}}$ for point (extended) sources \citep{Merloni12}. The horizontal lines in Figure \ref{fig:lc_BSG1} show the corresponding minimum detectable luminosity for the Galaxy and LMC. We note that the sensitivities are under the assumption of very little absorption, with column density $N_H=3\times 10^{20}\ {\rm cm^{-2}}$. This is too optimistic for Galactic sources, that are likely in the direction of the Galactic disk. As we see later, this assumption of small absorption is valid only for nearby ($<1$ kpc) ones. 

\section{Discussion}
\label{sec:discussion}
From Figure \ref{fig:lc_BSG1}, we obtain the observable duration of SNRs from failed SNe from BSGs, depending on the mass-loss rate $\dot{M}$. Table \ref{table:durations} shows the results for each BSG model, for a Galactic source 10 kpc away and a source in the LMC.

% \begin{table*}
% \centering
% \begin{tabular}{c|cccc}
% $\dot{M}$[$M_\odot\ {\rm yr^{-1}}$] & BSG-1, MW & BSG-1, LMC & BSG-2, MW & BSG-2, LMC\\ \hline
% $3\times 10^{-7}$ & $4.2\times 10^3$ & $2.9\times 10^3$  & $2.9\times 10^3$ & $1.5\times 10^3$ \\
% $1\times 10^{-6}$ & $2.6\times 10^4$ & $1.3\times 10^4$ & $1.8\times 10^4$ & $5.4\times 10^3$ \\
% $3\times 10^{-6}$ & $9.6\times 10^4$ & $6.9\times 10^4$ & $3.4\times 10^4$ & $8.2\times 10^3$ \\
% $6\times 10^{-6}$ & $8.8\times 10^4$ & $8.6 \times 10^4$& $1.2\times 10^3$ & $3\times 10^2$
% \end{tabular}
% \caption{The duration (in years) of the X-ray luminosity being above the sensitivity of eROSITA, at a distance of 10 kpc (corresponding to our Galaxy) and 50 kpc (LMC) neglecting absorption effects. If absorption is negligible, this number multiplied by our estimated event rate ($3\times 10^{-4}$ yr$^{-1}$ in our Galaxy, $10^{-4}$ yr$^{-1}$ in the LMC) corresponds to the expected number of detectable sources.}
% \label{table:durations}
% \end{table*}

\begin{table*}
\centering
\begin{tabular}{c|cccc}
$\dot{M}$[$M_\odot\ {\rm yr^{-1}}$] & BSG-1, MW & BSG-1, LMC & BSG-2, MW & BSG-2, LMC\\ \hline
$3\times 10^{-7}$ & $4.0\times 10^3$ & $2.9\times 10^3$  & $2.3\times 10^3$ & $1.3\times 10^3$ \\
$1\times 10^{-6}$ & $1.1\times 10^4$ & $4.5\times 10^3$ & $6.8\times 10^3$ & $1.7\times 10^3$ \\
$3\times 10^{-6}$ & $1.0\times 10^5$ & $2.0\times 10^4$ & $1.4\times 10^4$ & $1.5\times 10^3$ \\
$6\times 10^{-6}$ & $1.7\times 10^4$ & $1.6 \times 10^4$& $1.1\times 10^3$ & $2.9\times 10^2$
\end{tabular}
\caption{The duration (in years) of the X-ray luminosity being above the sensitivity of eROSITA, at a distance of 10 kpc (corresponding to our Galaxy) and 50 kpc (LMC) neglecting absorption effects. If absorption is negligible, this number multiplied by our estimated event rate ($3\times 10^{-4}$ yr$^{-1}$ in our Galaxy, $10^{-4}$ yr$^{-1}$ in the LMC) corresponds to the expected number of detectable sources.}
\label{table:durations}
\end{table*}

For Galactic sources, the maximum number of detectable sources, assuming an optimistic $\mathcal{R}_{\rm MW}=10^{-3}\ {\rm yr^{-1}}$ and negligible absorption, is $\approx 100$. However, an important caveat here is the assumption of low column density to estimate the sensitivity. Soft X-rays at around 1 keV start to be significantly absorbed when $N_H$ exceeds $10^{21}\ {\rm cm^{-2}}$. Assuming the HI gas in the ISM has number density of $10\ {\rm cm^{-3}}$ and filling fraction $3\%$ in the solar vicinity \citep{Tsuna18}, the observable sources in our Galaxy would be limited to within $\sim$ 1 kpc. The fraction of core-collapse occurring within 1 kpc from us is $\sim 1$\% \citep{Adams13}, which means that we actually can observe only 1\% of the $\leq 100$ sources. We thus conclude that Galactic sources would be difficult to detect, as we do not expect to have a source this nearby within the detectable period.

On the other hand, sources in the LMC can be more promising, as the typical foreground column density is low ($N_H<10^{21}\ {\rm cm^{-2}}$; \cite{Maggi16}). For sources in the LMC, the maximum time window of sources being brighter than the eROSITA sensitivity is $2\times 10^4 (2\times 10^3)$ yr for the BSG-1 (BSG-2) model. For our event rate of $10^{-4}$ yr$^{-1}$, there are 2 (0.2) detectable sources. However we note that (i) this is limited by flux rather than number of sources, and deeper surveys targeting the LMC can yield up to factor 5 more detections, (ii) fallback accretion, which may be common for BSGs \citep{Fernandez18}, may give additional energy to the ejecta and enhance X-ray emission.

In summary, the eROSITA all-sky survey with duration of 4 years may detect up to a few ``failed supernova remnants" from BSGs, likely those in the LMC. These sources are expected to have the following characteristics:
\begin{itemize}
    \item They have small size of order a few pcs, and slow shock velocities of $100$s of ${\rm km\ s^{-1}}$. 
    \item The thermal emission is mainly bright in soft X-rays, but dim in hard X-rays beyond a few keV.
    \item The spectra significantly lack signatures of heavy elements that should be created in normal SNRs, such as oxygen and iron. This is because the explosion is weak with negligible synthesis of heavy elements, and only the surface of the star compose the ejecta.
\end{itemize}
Detection of these sources enable observational studies of mass ejection upon the formation of BHs. While we assumed a simple spherical ejecta and CSM, there can be asymmetries. This can be probed by telescopes with much better angular resolution like Chandra \citep{Weisskopf02}, and may reveal some information of the progenitor.

Finally we briefly comment on the detectability of failed SNRs from other types of progenitors (RSGs and WRs). Although the rates of BH formation from RSGs and WRs may be higher than that of BSGs, we predict that these are more difficult to be detectable as SNRs. 

For RSGs, the ejecta velocity is slow, and the downstream temperature is a few eV assuming a strong shock. The cooling function is high enough that the shock becomes radiative. At the radiative limit the luminosity is
$
L\approx 2\pi r_{\rm sh}^2n_{\rm CSM}m_p v_{\rm sh}^3 
\sim 10L_\odot\left(v_{\rm sh}/{70\ {\rm km\  s^{-1}}}\right)^3 \left(\dot{M}/10^{-5}{\rm M_\odot yr^{-1}}\right)\left(v_w/10\ {\rm km\  s^{-1}}\right)^{-1}.
$
We thus predict a source with typical luminosity $\lesssim 10L_\odot$, but having a pc-scale thin-shell structure for an SNR of age $10^4$ yr. These sources are likely challenging to find by current optical/UV surveys. Perhaps the most promising strategy of probing failed SNe from RSGs is directly monitoring a large number of them, which was done in the past decade and identified a strong candidate \citep{Kochanek08,Gerke15,Adams17a}.

For WRs, the ejecta are light and thus decelerate very rapidly. For a typical WR wind of $\dot{M}=10^{-5}M_\odot\ {\rm yr}^{-1}$ and $v_w=10^3\ {\rm km\ s^{-1}}$, the Sedov timescale is only order 10 years. The ejecta velocity soon after becomes comparable to the $v_w$, after which the forward shock vanishes. 

\begin{ack}
The author thanks Koji Mori, Toshikazu Shigeyama, Kazumi Kashiyama and the referee for valuable comments, and thanks Kuniaki Masai for sharing his code for the X-ray spectra. This work is supported by the Advanced Leading Graduate Course for Photon Science (ALPS) at the University of Tokyo, and by JSPS KAKENHI Grant Number JP19J21578, MEXT, Japan.
\end{ack}

\bibliographystyle{apj} 
\bibliography{failedSNR}

\begin{thebibliography}{41}
\expandafter\ifx\csname natexlab\endcsname\relax\def\natexlab#1{#1}\fi

\bibitem[{{Abbott} {et~al.}(2019){Abbott}, {Abbott}, {Abbott}, {Abraham},
  {Acernese}, {Ackley}, {Adams}, {Adhikari}, {Adya}, {Affeldt}, {Agathos},
  {Agatsuma}, {Aggarwal}, {Aguiar}, {Aiello}, {Ain}, {Ajith}, {Allen},
  {Allocca}, {Aloy}, {Altin}, {Amato}, {Ananyeva}, {Anderson}, {Anderson},
  {Angelova}, {Antier}, {Appert}, {Arai}, {Araya}, {Areeda}, {Ar{\`e}ne},
  {Arnaud}, {Arun}, {Ascenzi}, {Ashton}, {Aston}, {Astone}, {Aubin}, {Aufmuth},
  {AultONeal}, {Austin}, {Avendano}, {Avila-Alvarez}, {Babak}, {Bacon},
  {Badaracco}, {Bader}, {Bae}, {Baker}, {Baldaccini}, {Ballardin}, {Ballmer},
  {Banagiri}, {Barayoga}, {Barclay}, {Barish}, {Barker}, {Barkett}, {Barnum},
  {Barone}, {Barr}, {Barsotti}, {Barsuglia}, {Barta}, {Bartlett}, {Bartos},
  {Bassiri}, {Basti}, {Bawaj}, {Bayley}, {Bazzan}, {B{\'e}csy}, {Bejger},
  {Belahcene}, {Bell}, {Beniwal}, {Berger}, {Bergmann}, {Bernuzzi}, {Bero},
  {Berry}, {Bersanetti}, {Bertolini}, {Betzwieser}, {Bhandare}, {Bidler},
  {Bilenko}, {Bilgili}, {Billingsley}, {Birch}, {Birney}, {Birnholtz},
  {Biscans}, {Biscoveanu}, {Bisht}, {Bitossi}, {Bizouard}, {Blackburn},
  {Blackman}, {Blair}, {Blair}, {Blair}, {Bloemen}, {Bode}, {Boer}, {Boetzel},
  {Bogaert}, {Bondu}, {Bonilla}, {Bonnand}, {Booker}, {Boom}, {Booth}, {Bork},
  {Boschi}, {Bose}, {Bossie}, {Bossilkov}, {Bosveld}, {Bouffanais}, {Bozzi},
  {Bradaschia}, {Brady}, {Bramley}, {Branchesi}, {Brau}, {Briant}, {Briggs},
  {Brighenti}, {Brillet}, {Brinkmann}, {Brisson}, {Brockill}, {Brooks},
  {Brown}, {Brunett}, {Buikema}, {Bulik}, {Bulten}, {Buonanno}, {Buskulic},
  {Bustamante Rosell}, {Buy}, {Byer}, {Cabero}, {Cadonati}, {Cagnoli},
  {Cahillane}, {Calder{\'o}n Bustillo}, {Callister}, {Calloni}, {Camp},
  {Campbell}, {Canepa}, {Cannon}, {Cao}, {Cao}, {Capocasa}, {Carbognani},
  {Caride}, {Carney}, {Carullo}, {Casanueva Diaz}, {Casentini}, {Caudill},
  {Cavagli{\`a}}, {Cavalier}, {Cavalieri}, {Cella}, {Cerd{\'a}-Dur{\'a}n},
  {Cerretani}, {Cesarini}, {Chaibi}, {Chakravarti}, {Chamberlin}, {Chan},
  {Chao}, {Charlton}, {Chase}, {Chassande-Mottin}, {Chatterjee}, {Chaturvedi},
  {Chatziioannou}, {Cheeseboro}, {Chen}, {Chen}, {Chen}, {Cheng}, {Cheong},
  {Chia}, {Chincarini}, {Chiummo}, {Cho}, {Cho}, {Cho}, {Christensen}, {Chu},
  {Chua}, {Chung}, {Chung}, {Ciani}, {Ciobanu}, {Ciolfi}, {Cipriano}, {Cirone},
  {Clara}, {Clark}, {Clearwater}, {Cleva}, {Cocchieri}, {Coccia}, {Cohadon},
  {Cohen}, {Colgan}, {Colleoni}, {Collette}, {Collins}, {Cominsky},
  {Constancio}, {Conti}, {Cooper}, {Corban}, {Corbitt}, {Cordero-Carri{\'o}n},
  {Corley}, {Cornish}, {Corsi}, {Cortese}, {Costa}, {Cotesta}, {Coughlin},
  {Coughlin}, {Coulon}, {Countryman}, {Couvares}, {Covas}, {Cowan}, {Coward},
  {Cowart}, {Coyne}, {Coyne}, {Creighton}, {Creighton}, {Cripe}, {Croquette},
  {Crowder}, {Cullen}, {Cumming}, {Cunningham}, {Cuoco}, {Canton}, {D{\'a}lya},
  {Danilishin}, {D'Antonio}, {Danzmann}, {Dasgupta}, {Da Silva Costa},
  {Datrier}, {Dattilo}, {Dave}, {Davier}, {Davis}, {Daw}, {DeBra},
  {Deenadayalan}, {Degallaix}, {De Laurentis}, {Del{\'e}glise}, {Del Pozzo},
  {DeMarchi}, {Demos}, {Dent}, {De Pietri}, {Derby}, {De Rosa}, {De Rossi},
  {DeSalvo}, {de Varona}, {Dhurandhar}, {D{\'\i}az}, {Dietrich}, {Di Fiore},
  {Di Giovanni}, {Di Girolamo}, {Di Lieto}, {Ding}, {Di Pace}, {Di Palma}, {Di
  Renzo}, {Dmitriev}, {Doctor}, {Donovan}, {Dooley}, {Doravari}, {Dorrington},
  {Downes}, {Drago}, {Driggers}, {Du}, {Ducoin}, {Dupej}, {Dwyer}, {Easter},
  {Edo}, {Edwards}, {Effler}, {Ehrens}, {Eichholz}, {Eikenberry}, {Eisenmann},
  {Eisenstein}, {Essick}, {Estelles}, {Estevez}, {Etienne}, {Etzel}, {Evans},
  {Evans}, {Fafone}, {Fair}, {Fairhurst}, {Fan}, {Farinon}, {Farr}, {Farr},
  {Fauchon-Jones}, {Favata}, {Fays}, {Fazio}, {Fee}, {Feicht}, {Fejer}, {Feng},
  {Fernandez-Galiana}, {Ferrante}, {Ferreira}, {Ferreira}, {Ferrini},
  {Fidecaro}, {Fiori}, {Fiorucci}, {Fishbach}, {Fisher}, {Fishner},
  {Fitz-Axen}, {Flaminio}, {Fletcher}, {Flynn}, {Fong}, {Font}, {Forsyth},
  {Fournier}, {Frasca}, {Frasconi}, {Frei}, {Freise}, {Frey}, {Frey},
  {Fritschel}, {Frolov}, {Fulda}, {Fyffe}, {Gabbard}, {Gadre}, {Gaebel},
  {Gair}, {Gammaitoni}, {Ganija}, {Gaonkar}, {Garcia},
  {Garc{\'\i}a-Quir{\'o}s}, {Garufi}, {Gateley}, {Gaudio}, {Gaur}, {Gayathri},
  {Gemme}, {Genin}, {Gennai}, {George}, {George}, {Gergely}, {Germain},
  {Ghonge}, {Ghosh}, {Ghosh}, {Ghosh}, {Giacomazzo}, {Giaime}, {Giardina},
  {Giazotto}, {Gill}, {Giordano}, {Glover}, {Godwin}, {Goetz}, {Goetz},
  {Goncharov}, {Gonz{\'a}lez}, {Gonzalez Castro}, {Gopakumar}, {Gorodetsky},
  {Gossan}, {Gosselin}, {Gouaty}, {Grado}, {Graef}, {Granata}, {Grant}, {Gras},
  {Grassia}, {Gray}, {Gray}, {Greco}, {Green}, {Green}, {Gretarsson}, {Groot},
  {Grote}, {Grunewald}, {Gruning}, {Guidi}, {Gulati}, {Guo}, {Gupta}, {Gupta},
  {Gustafson}, {Gustafson}, {Haegel}, {Halim}, {Hall}, {Hall}, {Hamilton},
  {Hammond}, {Haney}, {Hanke}, {Hanks}, {Hanna}, {Hannam}, {Hannuksela},
  {Hanson}, {Hardwick}, {Haris}, {Harms}, {Harry}, {Harry}, {Haster},
  {Haughian}, {Hayes}, {Healy}, {Heidmann}, {Heintze}, {Heitmann}, {Hello},
  {Hemming}, {Hendry}, {Heng}, {Hennig}, {Heptonstall}, {Hernandez Vivanco},
  {Heurs}, {Hild}, {Hinderer}, {Hoak}, {Hochheim}, {Hofman}, {Holgado},
  {Holland}, {Holt}, {Holz}, {Hopkins}, {Horst}, {Hough}, {Howell}, {Hoy},
  {Hreibi}, {Huang}, {Huerta}, {Huet}, {Hughey}, {Hulko}, {Husa}, {Huttner},
  {Huynh-Dinh}, {Idzkowski}, {Iess}, {Ingram}, {Inta}, {Intini}, {Irwin},
  {Isa}, {Isac}, {Isi}, {Iyer}, {Izumi}, {Jacqmin}, {Jadhav}, {Jani},
  {Janthalur}, {Jaranowski}, {Jenkins}, {Jiang}, {Johnson}, {Johnson-McDaniel},
  {Jones}, {Jones}, {Jones}, {Jonker}, {Ju}, {Junker}, {Kalaghatgi},
  {Kalogera}, {Kamai}, {Kandhasamy}, {Kang}, {Kanner}, {Kapadia}, {Karki},
  {Karvinen}, {Kashyap}, {Kasprzack}, {Katsanevas}, {Katsavounidis}, {Katzman},
  {Kaufer}, {Kawabe}, {Keerthana}, {K{\'e}f{\'e}lian}, {Keitel}, {Kennedy},
  {Key}, {Khalili}, {Khan}, {Khan}, {Khan}, {Khan}, {Khazanov}, {Khursheed},
  {Kijbunchoo}, {Kim}, {Kim}, {Kim}, {Kim}, {Kim}, {Kim}, {Kimball}, {King},
  {King}, {Kinley-Hanlon}, {Kirchhoff}, {Kissel}, {Kleybolte}, {Klika},
  {Klimenko}, {Knowles}, {Koch}, {Koehlenbeck}, {Koekoek}, {Koley},
  {Kondrashov}, {Kontos}, {Koper}, {Korobko}, {Korth}, {Kowalska}, {Kozak},
  {Kringel}, {Krishnendu}, {Kr{\'o}lak}, {Kuehn}, {Kumar}, {Kumar}, {Kumar},
  {Kumar}, {Kuo}, {Kutynia}, {Kwang}, {Lackey}, {Lai}, {Lam}, {Landry}, {Lane},
  {Lang}, {Lange}, {Lantz}, {Lanza}, {Lartaux-Vollard}, {Lasky}, {Laxen},
  {Lazzarini}, {Lazzaro}, {Leaci}, {Leavey}, {Lecoeuche}, {Lee}, {Lee}, {Lee},
  {Lee}, {Lee}, {Lee}, {Lehmann}, {Lenon}, {Leroy}, {Letendre}, {Levin}, {Li},
  {Li}, {Li}, {Li}, {Lin}, {Linde}, {Linker}, {Littenberg}, {Liu}, {Liu}, {Lo},
  {Lockerbie}, {London}, {Longo}, {Lorenzini}, {Loriette}, {Lormand},
  {Losurdo}, {Lough}, {Lousto}, {Lovelace}, {Lower}, {L{\"u}ck}, {Lumaca},
  {Lundgren}, {Lynch}, {Ma}, {Macas}, {Macfoy}, {MacInnis}, {Macleod},
  {Macquet}, {Maga{\~n}a-Sandoval}, {Maga{\~n}a Zertuche}, {Magee}, {Majorana},
  {Maksimovic}, {Malik}, {Man}, {Mandic}, {Mangano}, {Mansell}, {Manske},
  {Mantovani}, {Marchesoni}, {Marion}, {M{\'a}rka}, {M{\'a}rka}, {Markakis},
  {Markosyan}, {Markowitz}, {Maros}, {Marquina}, {Marsat}, {Martelli},
  {Martin}, {Martin}, {Martynov}, {Mason}, {Massera}, {Masserot}, {Massinger},
  {Masso-Reid}, {Mastrogiovanni}, {Matas}, {Matichard}, {Matone}, {Mavalvala},
  {Mazumder}, {McCann}, {McCarthy}, {McClelland}, {McCormick}, {McCuller},
  {McGuire}, {McIver}, {McManus}, {McRae}, {McWilliams}, {Meacher}, {Meadors},
  {Mehmet}, {Mehta}, {Meidam}, {Melatos}, {Mendell}, {Mercer}, {Mereni},
  {Merilh}, {Merzougui}, {Meshkov}, {Messenger}, {Messick}, {Metzdorff},
  {Meyers}, {Miao}, {Michel}, {Middleton}, {Mikhailov}, {Milano}, {Miller},
  {Miller}, {Millhouse}, {Mills}, {Milovich-Goff}, {Minazzoli}, {Minenkov},
  {Mishkin}, {Mishra}, {Mistry}, {Mitra}, {Mitrofanov}, {Mitselmakher},
  {Mittleman}, {Mo}, {Moffa}, {Mogushi}, {Mohapatra}, {Montani}, {Moore},
  {Moraru}, {Moreno}, {Morisaki}, {Mours}, {Mow-Lowry}, {Mukherjee},
  {Mukherjee}, {Mukherjee}, {Mukund}, {Mullavey}, {Munch}, {Mu{\~n}iz},
  {Muratore}, {Murray}, {Nagar}, {Nardecchia}, {Naticchioni}, {Nayak},
  {Neilson}, {Nelemans}, {Nelson}, {Nery}, {Neunzert}, {Ng}, {Ng}, {Nguyen},
  {Nichols}, {Nielsen}, {Nissanke}, {Nitz}, {Nocera}, {North}, {Nuttall},
  {Obergaulinger}, {Oberling}, {O'Brien}, {O'Dea}, {Ogin}, {Oh}, {Oh}, {Ohme},
  {Ohta}, {Okada}, {Oliver}, {Oppermann}, {Oram}, {O'Reilly}, {Ormiston},
  {Ortega}, {O'Shaughnessy}, {Ossokine}, {Ottaway}, {Overmier}, {Owen}, {Pace},
  {Pagano}, {Page}, {Pai}, {Pai}, {Palamos}, {Palashov}, {Palomba},
  {Pal-Singh}, {Pan}, {Pang}, {Pang}, {Pankow}, {Pannarale}, {Pant},
  {Paoletti}, {Paoli}, {Papa}, {Parida}, {Parker}, {Pascucci}, {Pasqualetti},
  {Passaquieti}, {Passuello}, {Patil}, {Patricelli}, {Pearlstone}, {Pedersen},
  {Pedraza}, {Pedurand}, {Pele}, {Penn}, {Perego}, {Perez}, {Perreca},
  {Pfeiffer}, {Phelps}, {Phukon}, {Piccinni}, {Pichot}, {Piergiovanni},
  {Pillant}, {Pinard}, {Pirello}, {Pitkin}, {Poggiani}, {Pong}, {Ponrathnam},
  {Popolizio}, {Porter}, {Powell}, {Prajapati}, {Prasad}, {Prasai}, {Prasanna},
  {Pratten}, {Prestegard}, {Privitera}, {Prodi}, {Prokhorov}, {Puncken},
  {Punturo}, {Puppo}, {P{\"u}rrer}, {Qi}, {Quetschke}, {Quinonez}, {Quintero},
  {Quitzow-James}, {Raab}, {Radkins}, {Radulescu}, {Raffai}, {Raja}, {Rajan},
  {Rajbhandari}, {Rakhmanov}, {Ramirez}, {Ramos-Buades}, {Rana}, {Rao},
  {Rapagnani}, {Raymond}, {Razzano}, {Read}, {Regimbau}, {Rei}, {Reid},
  {Reitze}, {Ren}, {Ricci}, {Richardson}, {Richardson}, {Ricker},
  {Riemenschneider}, {Riles}, {Rizzo}, {Robertson}, {Robie}, {Robinet},
  {Rocchi}, {Rolland}, {Rollins}, {Roma}, {Romanelli}, {Romano}, {Romel},
  {Romie}, {Rose}, {Rosi{\'n}ska}, {Rosofsky}, {Ross}, {Rowan}, {R{\"u}diger},
  {Ruggi}, {Rutins}, {Ryan}, {Sachdev}, {Sadecki}, {Sakellariadou}, {Salafia},
  {Salconi}, {Saleem}, {Salemi}, {Samajdar}, {Sammut}, {Sanchez}, {Sanchez},
  {Sanchis-Gual}, {Sandberg}, {Sanders}, {Santiago}, {Sarin}, {Sassolas},
  {Sathyaprakash}, {Saulson}, {Sauter}, {Savage}, {Schale}, {Scheel},
  {Scheuer}, {Schmidt}, {Schnabel}, {Schofield}, {Sch{\"o}nbeck}, {Schreiber},
  {Schulte}, {Schutz}, {Schwalbe}, {Scott}, {Scott}, {Seidel}, {Sellers},
  {Sengupta}, {Sennett}, {Sentenac}, {Sequino}, {Sergeev}, {Setyawati},
  {Shaddock}, {Shaffer}, {Shahriar}, {Shaner}, {Shao}, {Sharma}, {Shawhan},
  {Shen}, {Shink}, {Shoemaker}, {Shoemaker}, {ShyamSundar}, {Siellez},
  {Sieniawska}, {Sigg}, {Silva}, {Singer}, {Singh}, {Singhal}, {Sintes},
  {Sitmukhambetov}, {Skliris}, {Slagmolen}, {Slaven-Blair}, {Smith}, {Smith},
  {Somala}, {Son}, {Sorazu}, {Sorrentino}, {Souradeep}, {Sowell}, {Spencer},
  {Srivastava}, {Srivastava}, {Staats}, {Stachie}, {Standke}, {Steer},
  {Steinke}, {Steinlechner}, {Steinlechner}, {Steinmeyer}, {Stevenson},
  {Stocks}, {Stone}, {Stops}, {Strain}, {Stratta}, {Strigin}, {Strunk},
  {Sturani}, {Stuver}, {Sudhir}, {Summerscales}, {Sun}, {Sunil}, {Suresh},
  {Sutton}, {Swinkels}, {Szczepa{\'n}czyk}, {Tacca}, {Tait}, {Talbot},
  {Talukder}, {Tanner}, {T{\'a}pai}, {Taracchini}, {Tasson}, {Taylor}, {Thies},
  {Thomas}, {Thomas}, {Thondapu}, {Thorne}, {Thrane}, {Tiwari}, {Tiwari},
  {Tiwari}, {Toland}, {Tonelli}, {Tornasi}, {Torres-Forn{\'e}}, {Torrie},
  {T{\"o}yr{\"a}}, {Travasso}, {Traylor}, {Tringali}, {Trovato}, {Trozzo},
  {Trudeau}, {Tsang}, {Tse}, {Tso}, {Tsukada}, {Tsuna}, {Tuyenbayev}, {Ueno},
  {Ugolini}, {Unnikrishnan}, {Urban}, {Usman}, {Vahlbruch}, {Vajente},
  {Valdes}, {van Bakel}, {van Beuzekom}, {van den Brand}, {Van Den Broeck},
  {Vander-Hyde}, {van Heijningen}, {van der Schaaf}, {van Veggel}, {Vardaro},
  {Varma}, {Vass}, {Vas{\'u}th}, {Vecchio}, {Vedovato}, {Veitch}, {Veitch},
  {Venkateswara}, {Venugopalan}, {Verkindt}, {Vetrano}, {Vicer{\'e}}, {Viets},
  {Vine}, {Vinet}, {Vitale}, {Vo}, {Vocca}, {Vorvick}, {Vyatchanin}, {Wade},
  {Wade}, {Wade}, {Walet}, {Walker}, {Wallace}, {Walsh}, {Wang}, {Wang},
  {Wang}, {Wang}, {Wang}, {Ward}, {Warden}, {Warner}, {Was}, {Watchi},
  {Weaver}, {Wei}, {Weinert}, {Weinstein}, {Weiss}, {Wellmann}, {Wen},
  {Wessel}, {We{\ss}els}, {Westhouse}, {Wette}, {Whelan}, {White}, {Whiting},
  {Whittle}, {Wilken}, {Williams}, {Williamson}, {Willis}, {Willke}, {Wimmer},
  {Winkler}, {Wipf}, {Wittel}, {Woan}, {Woehler}, {Wofford}, {Worden},
  {Wright}, {Wu}, {Wysocki}, {Xiao}, {Yamamoto}, {Yancey}, {Yang}, {Yap},
  {Yazback}, {Yeeles}, {Yu}, {Yu}, {Yuen}, {Yvert}, {Zadro{\.Z}ny}, {Zanolin},
  {Zappa}, {Zelenova}, {Zendri}, {Zevin}, {Zhang}, {Zhang}, {Zhang}, {Zhao},
  {Zhou}, {Zhou}, {Zhu}, {Zimmerman}, {Zlochower}, {Zucker}, {Zweizig}, {LIGO
  Scientific Collaboration}, \& {Virgo Collaboration}}]{Abbott19}
{Abbott}, B.~P., et~al. 2019, Physical Review X, 9, 031040

\bibitem[{{Abbott} {et~al.}(2020){Abbott}, {Abbott}, {Abraham}, {Acernese},
  {Ackley}, {Adams}, {Adams}, {Adhikari}, {Adya}, {Affeldt}, {Agathos},
  {Agatsuma}, {Aggarwal}, {Aguiar}, {Aiello}, {Ain}, {Ajith}, {Akcay}, {Allen},
  {Allocca}, {Altin}, {Amato}, {Anand}, {Ananyeva}, {Anderson}, {Anderson},
  {Angelova}, {Ansoldi}, {Antelis}, {Antier}, {Appert}, {Arai}, {Araya},
  {Areeda}, {Ar{\`e}ne}, {Arnaud}, {Aronson}, {Arun}, {Asali}, {Ascenzi},
  {Ashton}, {Aston}, {Astone}, {Aubin}, {Aufmuth}, {AultONeal}, {Austin},
  {Avendano}, {Babak}, {Badaracco}, {Bader}, {Bae}, {Baer}, {Bagnasco},
  {Baird}, {Ball}, {Ballardin}, {Ballmer}, {Bals}, {Balsamo}, {Baltus},
  {Banagiri}, {Bankar}, {Bankar}, {Barayoga}, {Barbieri}, {Barish}, {Barker},
  {Barneo}, {Barnum}, {Barone}, {Barr}, {Barsotti}, {Barsuglia}, {Barta},
  {Bartlett}, {Bartos}, {Bassiri}, {Basti}, {Bawaj}, {Bayley}, {Bazzan},
  {Becher}, {B{\'e}csy}, {Bedakihale}, {Bejger}, {Belahcene}, {Beniwal},
  {Benjamin}, {Bennett}, {Bentley}, {Bergamin}, {Berger}, {Bergmann},
  {Bernuzzi}, {Berry}, {Bersanetti}, {Bertolini}, {Betzwieser}, {Bhandare},
  {Bhandari}, {Bhattacharjee}, {Bidler}, {Bilenko}, {Billingsley}, {Birney},
  {Birnholtz}, {Biscans}, {Bischi}, {Biscoveanu}, {Bisht}, {Bitossi},
  {Bizouard}, {Blackburn}, {Blackman}, {Blair}, {Blair}, {Blair}, {Blanch},
  {Bobba}, {Bode}, {Boer}, {Boetzel}, {Bogaert}, {Boldrini}, {Bondu},
  {Bonnand}, {Bonilla}, {Booker}, {Boom}, {Bork}, {Boschi}, {Bose},
  {Bossilkov}, {Boudart}, {Bouffanais}, {Bozzi}, {Bradaschia}, {Brady},
  {Bramley}, {Branchesi}, {Brau}, {Breschi}, {Briant}, {Briggs}, {Brighenti},
  {Brillet}, {Brinkmann}, {Brockill}, {Brooks}, {Brooks}, {Brown}, {Brunett},
  {Bruno}, {Bruntz}, {Buikema}, {Bulik}, {Bulten}, {Buonanno}, {Buscicchio},
  {Buskulic}, {Byer}, {Cabero}, {Cadonati}, {Caesar}, {Cagnoli}, {Cahillane},
  {Calder{\'o}n Bustillo}, {Callaghan}, {Callister}, {Calloni}, {Camp},
  {Canepa}, {Cannon}, {Cao}, {Cao}, {Carapella}, {Carbognani}, {Carney},
  {Carpinelli}, {Carullo}, {Carver}, {Casanueva Diaz}, {Casentini}, {Caudill},
  {Cavagli{\`a}}, {Cavalier}, {Cavalieri}, {Cella}, {Cerd{\'a}-Dur{\'a}n},
  {Cesarini}, {Chaibi}, {Chakravarti}, {Chan}, {Chan}, {Chandra}, {Chanial},
  {Chao}, {Charlton}, {Chase}, {Chassande-Mottin}, {Chatterjee},
  {Chattopadhyay}, {Chaturvedi}, {Chatziioannou}, {Chen}, {Chen}, {Chen},
  {Chen}, {Cheng}, {Cheong}, {Chia}, {Chiadini}, {Chierici}, {Chincarini},
  {Chiummo}, {Cho}, {Cho}, {Cho}, {Choate}, {Christensen}, {Chu}, {Chua},
  {Chung}, {Chung}, {Ciani}, {Ciecielag}, {Cie{\'s}lar}, {Cifaldi}, {Ciobanu},
  {Ciolfi}, {Cipriano}, {Cirone}, {Clara}, {Clark}, {Clark}, {Clarke},
  {Clearwater}, {Clesse}, {Cleva}, {Coccia}, {Cohadon}, {Cohen}, {Colleoni},
  {Collette}, {Collins}, {Colpi}, {Constancio}, {Conti}, {Cooper}, {Corban},
  {Corbitt}, {Cordero-Carri{\'o}n}, {Corezzi}, {Corley}, {Cornish}, {Corre},
  {Corsi}, {Cortese}, {Costa}, {Cotesta}, {Coughlin}, {Coughlin}, {Coulon},
  {Countryman}, {Cousins}, {Couvares}, {Covas}, {Coward}, {Cowart}, {Coyne},
  {Coyne}, {Creighton}, {Creighton}, {Croquette}, {Crowder}, {Cudell},
  {Cullen}, {Cumming}, {Cummings}, {Cunningham}, {Cuoco}, {Curylo}, {Dal
  Canton}, {D{\'a}lya}, {Dana}, {DaneshgaranBajastani}, {D'Angelo}, {Danila},
  {Danilishin}, {D'Antonio}, {Danzmann}, {Darsow-Fromm}, {Dasgupta}, {Datrier},
  {Dattilo}, {Dave}, {Davier}, {Davies}, {Davis}, {Daw}, {Dean}, {DeBra},
  {Deenadayalan}, {Degallaix}, {De Laurentis}, {Del{\'e}glise}, {Del Favero},
  {De Lillo}, {De Lillo}, {Del Pozzo}, {DeMarchi}, {De Matteis}, {D'Emilio},
  {Demos}, {Denker}, {Dent}, {Depasse}, {De Pietri}, {De Rosa}, {De Rossi},
  {DeSalvo}, {de Varona}, {Dhurandhar}, {D{\'\i}az}, {Diaz-Ortiz}, {Didio},
  {Dietrich}, {Di Fiore}, {DiFronzo}, {Di Giorgio}, {Di Giovanni}, {Di
  Giovanni}, {Di Girolamo}, {Di Lieto}, {Ding}, {Di Pace}, {Di Palma}, {Di
  Renzo}, {Divakarla}, {Dmitriev}, {Doctor}, {D'Onofrio}, {Donovan}, {Dooley},
  {Doravari}, {Dorrington}, {Downes}, {Drago}, {Driggers}, {Du}, {Ducoin},
  {Dupej}, {Durante}, {D'Urso}, {Duverne}, {Dwyer}, {Easter}, {Eddolls},
  {Edelman}, {Edo}, {Edy}, {Effler}, {Eichholz}, {Eikenberry}, {Eisenmann},
  {Eisenstein}, {Ejlli}, {Errico}, {Essick}, {Estell{\'e}s}, {Estevez},
  {Etienne}, {Etzel}, {Evans}, {Evans}, {Ewing}, {Fafone}, {Fair}, {Fairhurst},
  {Fan}, {Farah}, {Farinon}, {Farr}, {Farr}, {Fauchon-Jones}, {Favata}, {Fays},
  {Fazio}, {Feicht}, {Fejer}, {Feng}, {Fenyvesi}, {Ferguson},
  {Fernandez-Galiana}, {Ferrante}, {Ferreira}, {Fidecaro}, {Figura}, {Fiori},
  {Fiorucci}, {Fishbach}, {Fisher}, {Fishner}, {Fittipaldi}, {Fitz-Axen},
  {Fiumara}, {Flaminio}, {Floden}, {Flynn}, {Fong}, {Font}, {Forsyth},
  {Fournier}, {Frasca}, {Frasconi}, {Frei}, {Freise}, {Frey}, {Frey},
  {Fritschel}, {Frolov}, {Fronz{\'e}}, {Fulda}, {Fyffe}, {Gabbard}, {Gadre},
  {Gaebel}, {Gair}, {Gais}, {Galaudage}, {Gamba}, {Ganapathy}, {Ganguly},
  {Gaonkar}, {Garaventa}, {Garc{\'\i}a-Quir{\'o}s}, {Garufi}, {Gateley},
  {Gaudio}, {Gayathri}, {Gemme}, {Gennai}, {George}, {George}, {George},
  {Gergely}, {Ghonge}, {Ghosh}, {Ghosh}, {Ghosh}, {Giacomazzo}, {Giacoppo},
  {Giaime}, {Giardina}, {Gibson}, {Gier}, {Gill}, {Giri}, {Glanzer}, {Gleckl},
  {Godwin}, {Goetz}, {Goetz}, {Gohlke}, {Goncharov}, {Gonz{\'a}lez},
  {Gopakumar}, {Gossan}, {Gosselin}, {Gouaty}, {Grace}, {Grado}, {Granata},
  {Granata}, {Grant}, {Gras}, {Grassia}, {Gray}, {Gray}, {Greco}, {Green},
  {Green}, {Gretarsson}, {Griggs}, {Grignani}, {Grimaldi}, {Grimes}, {Grimm},
  {Grote}, {Grunewald}, {Gruning}, {Guerrero}, {Guidi}, {Guimaraes},
  {Guix{\'e}}, {Gulati}, {Guo}, {Gupta}, {Gupta}, {Gupta}, {Gustafson},
  {Gustafson}, {Guzman}, {Haegel}, {Halim}, {Hall}, {Hamilton}, {Hammond},
  {Haney}, {Hanke}, {Hanks}, {Hanna}, {Hannam}, {Hannuksela}, {Hannuksela},
  {Hansen}, {Hansen}, {Hanson}, {Harder}, {Hardwick}, {Haris}, {Harms},
  {Harry}, {Harry}, {Hartwig}, {Hasskew}, {Haster}, {Haughian}, {Hayes},
  {Healy}, {Heidmann}, {Heintze}, {Heinze}, {Heinzel}, {Heitmann}, {Hellman},
  {Hello}, {Helmling-Cornell}, {Hemming}, {Hendry}, {Heng}, {Hennes}, {Hennig},
  {Hennig}, {Hernandez Vivanco}, {Heurs}, {Hild}, {Hill}, {Hines}, {Hochheim},
  {Hofgard}, {Hofman}, {Hohmann}, {Holgado}, {Holland}, {Hollows}, {Holmes},
  {Holt}, {Holz}, {Hopkins}, {Horst}, {Hough}, {Howell}, {Hoy}, {Hoyland},
  {Huang}, {H{\"u}bner}, {Huddart}, {Huerta}, {Hughey}, {Hui}, {Husa},
  {Huttner}, {Hutzler}, {Huxford}, {Huynh-Dinh}, {Idzkowski}, {Iess},
  {Imperato}, {Inchauspe}, {Ingram}, {Intini}, {Isi}, {Iyer},
  {JaberianHamedan}, {Jacqmin}, {Jadhav}, {Jadhav}, {James}, {Jani},
  {Janssens}, {Janthalur}, {Jaranowski}, {Jariwala}, {Jaume}, {Jenkins},
  {Jeunon}, {Jiang}, {Johns}, {Johnson-McDaniel}, {Jones}, {Jones}, {Jones},
  {Jones}, {Jones}, {Jonker}, {Ju}, {Junker}, {Kalaghatgi}, {Kalogera},
  {Kamai}, {Kandhasamy}, {Kang}, {Kanner}, {Kapadia}, {Kapasi}, {Karathanasis},
  {Karki}, {Kashyap}, {Kasprzack}, {Kastaun}, {Katsanevas}, {Katsavounidis},
  {Katzman}, {Kawabe}, {K{\'e}f{\'e}lian}, {Keitel}, {Key}, {Khadka},
  {Khalili}, {Khan}, {Khan}, {Khazanov}, {Khetan}, {Khursheed}, {Kijbunchoo},
  {Kim}, {Kim}, {Kim}, {Kim}, {Kim}, {Kim}, {Kimball}, {King}, {Kinley-Hanlon},
  {Kirchhoff}, {Kissel}, {Kleybolte}, {Klimenko}, {Knowles}, {Knyazev}, {Koch},
  {Koehlenbeck}, {Koekoek}, {Koley}, {Kolstein}, {Komori}, {Kondrashov},
  {Kontos}, {Koper}, {Korobko}, {Korth}, {Kovalam}, {Kozak}, {Kr{\"a}mer},
  {Kringel}, {Krishnendu}, {Kr{\'o}lak}, {Kuehn}, {Kumar}, {Kumar}, {Kumar},
  {Kumar}, {Kuns}, {Kwang}, {Lackey}, {Laghi}, {Lalande}, {Lam}, {Lamberts},
  {Landry}, {Lane}, {Lang}, {Lange}, {Lantz}, {Lanza}, {La Rosa},
  {Lartaux-Vollard}, {Lasky}, {Laxen}, {Lazzarini}, {Lazzaro}, {Leaci},
  {Leavey}, {Lecoeuche}, {Lee}, {Lee}, {Lee}, {Lee}, {Lehmann}, {Leon},
  {Leroy}, {Letendre}, {Levin}, {Li}, {Li}, {Li}, {Li}, {Li}, {Linde},
  {Linker}, {Linley}, {Littenberg}, {Liu}, {Liu}, {Llorens-Monteagudo}, {Lo},
  {Lockwood}, {London}, {Longo}, {Lorenzini}, {Loriette}, {Lormand}, {Losurdo},
  {Lough}, {Lousto}, {Lovelace}, {L{\"u}ck}, {Lumaca}, {Lundgren}, {Ma},
  {Macas}, {MacInnis}, {Macleod}, {MacMillan}, {Macquet}, {Maga{\~n}a
  Hernandez}, {Maga{\~n}a-Sandoval}, {Magazz{\`u}}, {Magee}, {Majorana},
  {Maksimovic}, {Maliakal}, {Malik}, {Man}, {Mandic}, {Mangano}, {Mansell},
  {Manske}, {Mantovani}, {Mapelli}, {Marchesoni}, {Marion}, {M{\'a}rka},
  {M{\'a}rka}, {Markakis}, {Markosyan}, {Markowitz}, {Maros}, {Marquina},
  {Marsat}, {Martelli}, {Martin}, {Martin}, {Martinez}, {Martinez}, {Martynov},
  {Masalehdan}, {Mason}, {Massera}, {Masserot}, {Massinger}, {Masso-Reid},
  {Mastrogiovanni}, {Matas}, {Mateu-Lucena}, {Matichard}, {Matiushechkina},
  {Mavalvala}, {Maynard}, {McCann}, {McCarthy}, {McClelland}, {McCormick},
  {McCuller}, {McGuire}, {McIsaac}, {McIver}, {McManus}, {McRae}, {McWilliams},
  {Meacher}, {Meadors}, {Mehmet}, {Mehta}, {Melatos}, {Melchor}, {Mendell},
  {Menendez-Vazquez}, {Mercer}, {Mereni}, {Merfeld}, {Merilh}, {Merritt},
  {Merzougui}, {Meshkov}, {Messenger}, {Messick}, {Metzdorff}, {Meyers},
  {Meylahn}, {Mhaske}, {Miani}, {Miao}, {Michaloliakos}, {Michel}, {Middleton},
  {Milano}, {Miller}, {Millhouse}, {Mills}, {Milotti}, {Milovich-Goff},
  {Minazzoli}, {Minenkov}, {Mir}, {Mishkin}, {Mishra}, {Mistry}, {Mitra},
  {Mitrofanov}, {Mitselmakher}, {Mittleman}, {Mo}, {Mogushi}, {Mohapatra},
  {Mohite}, {Molina}, {Molina-Ruiz}, {Mondin}, {Montani}, {Moore}, {Moraru},
  {Morawski}, {Moreno}, {Morisaki}, {Mours}, {Mow-Lowry}, {Mozzon},
  {Muciaccia}, {Mukherjee}, {Mukherjee}, {Mukherjee}, {Mukherjee}, {Mukund},
  {Mullavey}, {Munch}, {Mu{\~n}iz}, {Murray}, {Nadji}, {Nagar}, {Nardecchia},
  {Naticchioni}, {Nayak}, {Neil}, {Neilson}, {Nelemans}, {Nelson}, {Nery},
  {Neunzert}, {Nitz}, {Ng}, {Ng}, {Nguyen}, {Nguyen}, {Nguyen}, {Nichols},
  {Nissanke}, {Nocera}, {Noh}, {North}, {Nothard}, {Nuttall}, {Oberling},
  {O'Brien}, {O'Dell}, {Oganesyan}, {Ogin}, {Oh}, {Oh}, {Ohme}, {Ohta},
  {Okada}, {Olivetto}, {Oppermann}, {Oram}, {O'Reilly}, {Ormiston}, {Ortega},
  {O'Shaughnessy}, {Ossokine}, {Osthelder}, {Ottaway}, {Overmier}, {Owen},
  {Pace}, {Pagano}, {Page}, {Pagliaroli}, {Pai}, {Pai}, {Palamos}, {Palashov},
  {Palomba}, {Pan}, {Panda}, {Pang}, {Pankow}, {Pannarale}, {Pant}, {Paoletti},
  {Paoli}, {Paolone}, {Parker}, {Pascucci}, {Pasqualetti}, {Passaquieti},
  {Passuello}, {Patel}, {Patricelli}, {Payne}, {Pechsiri}, {Pedraza},
  {Pegoraro}, {Pele}, {Penn}, {Perego}, {Perez}, {P{\'e}rigois}, {Perreca},
  {Perri{\`e}s}, {Petermann}, {Petterson}, {Pfeiffer}, {Pham}, {Phukon},
  {Piccinni}, {Pichot}, {Piendibene}, {Piergiovanni}, {Pierini}, {Pierro},
  {Pillant}, {Pilo}, {Pinard}, {Pinto}, {Piotrzkowski}, {Pirello}, {Pitkin},
  {Placidi}, {Plastino}, {Pluchar}, {Poggiani}, {Polini}, {Pong}, {Ponrathnam},
  {Popolizio}, {Porter}, {Poverman}, {Powell}, {Pracchia}, {Prajapati},
  {Prasai}, {Prasanna}, {Pratten}, {Prestegard}, {Principe}, {Prodi},
  {Prokhorov}, {Prosposito}, {Prudenzi}, {Puecher}, {Punturo}, {Puosi},
  {Puppo}, {P{\"u}rrer}, {Qi}, {Quetschke}, {Quinonez}, {Quitzow-James},
  {Raab}, {Raaijmakers}, {Radkins}, {Radulesco}, {Raffai}, {Rafferty}, {Rail},
  {Raja}, {Rajan}, {Rajbhandari}, {Rakhmanov}, {Ramirez}, {Ramirez},
  {Ramos-Buades}, {Rana}, {Rao}, {Rapagnani}, {Rapol}, {Ratto}, {Raymond},
  {Razzano}, {Read}, {Regimbau}, {Rei}, {Reid}, {Reitze}, {Rettegno}, {Ricci},
  {Richardson}, {Richardson}, {Richardson}, {Ricker}, {Riemenschneider},
  {Riles}, {Rizzo}, {Robertson}, {Robinet}, {Rocchi}, {Rocha}, {Rodriguez},
  {Rodriguez-Soto}, {Rolland}, {Rollins}, {Roma}, {Romanelli}, {Romano},
  {Romel}, {Romero}, {Romero-Shaw}, {Romie}, {Ronchini}, {Rose}, {Rose},
  {Rose}, {Rosell}, {Rosi{\'n}ska}, {Rosofsky}, {Ross}, {Rowan}, {Rowlinson},
  {Roy}, {Roy}, {Ruggi}, {Ryan}, {Sachdev}, {Sadecki}, {Sadiq},
  {Sakellariadou}, {Salafia}, {Salconi}, {Saleem}, {Samajdar}, {Sanchez},
  {Sanchez}, {Sanchez}, {Sanchis-Gual}, {Sanders}, {Sandles}, {Santiago},
  {Santos}, {Saravanan}, {Sarin}, {Sassolas}, {Sathyaprakash}, {Sauter},
  {Savage}, {Savant}, {Sawant}, {Sayah}, {Schaetzl}, {Schale}, {Scheel},
  {Scheuer}, {Schindler-Tyka}, {Schmidt}, {Schnabel}, {Schofield},
  {Sch{\"o}nbeck}, {Schreiber}, {Schulte}, {Schutz}, {Schwarm}, {Schwartz},
  {Scott}, {Scott}, {Seglar-Arroyo}, {Seidel}, {Sellers}, {Sengupta},
  {Sennett}, {Sentenac}, {Sequino}, {Sergeev}, {Setyawati}, {Shaffer},
  {Shahriar}, {Sharifi}, {Sharma}, {Sharma}, {Shawhan}, {Shen}, {Shikauchi},
  {Shink}, {Shoemaker}, {Shoemaker}, {Shukla}, {ShyamSundar}, {Sieniawska},
  {Sigg}, {Singer}, {Singh}, {Singh}, {Singha}, {Singhal}, {Sintes}, {Sipala},
  {Skliris}, {Slagmolen}, {Slaven-Blair}, {Smetana}, {Smith}, {Smith},
  {Somala}, {Son}, {Soni}, {Soni}, {Sorazu}, {Sordini}, {Sorrentino},
  {Sorrentino}, {Soulard}, {Souradeep}, {Sowell}, {Spencer}, {Spera},
  {Srivastava}, {Srivastava}, {Staats}, {Stachie}, {Steer}, {Steinhoff},
  {Steinke}, {Steinlechner}, {Steinlechner}, {Steinmeyer}, {Stevenson},
  {Stolle-McAllister}, {Stops}, {Stover}, {Strain}, {Stratta}, {Strunk},
  {Sturani}, {Stuver}, {S{\"u}dbeck}, {Sudhagar}, {Sudhir}, {Suh},
  {Summerscales}, {Sun}, {Sun}, {Sunil}, {Sur}, {Suresh}, {Sutton}, {Swinkels},
  {Szczepa{\'n}czyk}, {Tacca}, {Tait}, {Talbot}, {Tanasijczuk}, {Tanner},
  {Tao}, {Tapia}, {Tapia San Martin}, {Tasson}, {Taylor}, {Tenorio},
  {Terkowski}, {Thirugnanasambandam}, {Thomas}, {Thomas}, {Thomas}, {Thompson},
  {Thondapu}, {Thorne}, {Thrane}, {Tiwari}, {Tiwari}, {Tiwari}, {Toland},
  {Tolley}, {Tonelli}, {Tornasi}, {Torres-Forn{\'e}}, {Torrie}, {Melo},
  {T{\"o}yr{\"a}}, {Tran}, {Trapananti}, {Travasso}, {Traylor}, {Tringali},
  {Tripathee}, {Trovato}, {Trudeau}, {Tsai}, {Tsang}, {Tse}, {Tso}, {Tsukada},
  {Tsuna}, {Tsutsui}, {Turconi}, {Ubhi}, {Udall}, {Ueno}, {Ugolini},
  {Unnikrishnan}, {Urban}, {Usman}, {Utina}, {Vahlbruch}, {Vajente}, {Vajpeyi},
  {Valdes}, {Valentini}, {Valsan}, {van Bakel}, {van Beuzekom}, {van den
  Brand}, {Van Den Broeck}, {Vander-Hyde}, {van der Schaaf}, {van Heijningen},
  {Vardaro}, {Vargas}, {Varma}, {Vass}, {Vas{\'u}th}, {Vecchio}, {Vedovato},
  {Veitch}, {Veitch}, {Venkateswara}, {Venneberg}, {Venugopalan}, {Verkindt},
  {Verma}, {Veske}, {Vetrano}, {Vicer{\'e}}, {Viets}, {Vijaykumar},
  {Villa-Ortega}, {Vinet}, {Vitale}, {Vo}, {Vocca}, {Vorvick}, {Vyatchanin},
  {Wade}, {Wade}, {Wade}, {Walet}, {Walker}, {Wallace}, {Wallace}, {Walsh},
  {Wang}, {Wang}, {Wang}, {Wang}, {Ward}, {Warner}, {Was}, {Washington},
  {Watchi}, {Weaver}, {Wei}, {Weinert}, {Weinstein}, {Weiss}, {Wellmann},
  {Wen}, {We{\ss}els}, {Westhouse}, {Wette}, {Whelan}, {White}, {White},
  {Whiting}, {Whittle}, {Wilken}, {Williams}, {Williams}, {Williamson},
  {Willis}, {Willke}, {Wilson}, {Wimmer}, {Winkler}, {Wipf}, {Woan}, {Woehler},
  {Wofford}, {Wong}, {Wrangel}, {Wright}, {Wu}, {Wysocki}, {Xiao}, {Yamamoto},
  {Yang}, {Yang}, {Yang}, {Yap}, {Yeeles}, {Yoon}, {Yu}, {Yu}, {Yuen},
  {Zadro{\.z}ny}, {Zanolin}, {Zelenova}, {Zendri}, {Zevin}, {Zhang}, {Zhang},
  {Zhang}, {Zhang}, {Zhao}, {Zhao}, {Zheng}, {Zhou}, {Zhou}, {Zhu},
  {Zimmerman}, {Zlochower}, {Zucker}, \& {Zweizig}}]{Abbott20}
{Abbott}, R., et~al. 2020, arXiv e-prints, arXiv:2010.14527

\bibitem[{{Adams} {et~al.}(2013){Adams}, {Kochanek}, {Beacom}, {Vagins}, \&
  {Stanek}}]{Adams13}
{Adams}, S.~M., {Kochanek}, C.~S., {Beacom}, J.~F., {Vagins}, M.~R., \&
  {Stanek}, K.~Z. 2013, \apj, 778, 164

\bibitem[{{Adams} {et~al.}(2017{\natexlab{a}}){Adams}, {Kochanek}, {Gerke}, \&
  {Stanek}}]{Adams17b}
{Adams}, S.~M., {Kochanek}, C.~S., {Gerke}, J.~R., \& {Stanek}, K.~Z.
  2017{\natexlab{a}}, \mnras, 469, 1445

\bibitem[{{Adams} {et~al.}(2017{\natexlab{b}}){Adams}, {Kochanek}, {Gerke},
  {Stanek}, \& {Dai}}]{Adams17a}
{Adams}, S.~M., {Kochanek}, C.~S., {Gerke}, J.~R., {Stanek}, K.~Z., \& {Dai},
  X. 2017{\natexlab{b}}, \mnras, 468, 4968

\bibitem[{{Castor} {et~al.}(1975){Castor}, {McCray}, \& {Weaver}}]{Castor75}
{Castor}, J., {McCray}, R., \& {Weaver}, R. 1975, \apjl, 200, L107

\bibitem[{{Corral-Santana} {et~al.}(2016){Corral-Santana}, {Casares},
  {Mu{\~n}oz-Darias}, {Bauer}, {Mart{\'\i}nez-Pais}, \& {Russell}}]{Corral16}
{Corral-Santana}, J.~M., {Casares}, J., {Mu{\~n}oz-Darias}, T., {Bauer}, F.~E.,
  {Mart{\'\i}nez-Pais}, I.~G., \& {Russell}, D.~M. 2016, \aap, 587, A61

\bibitem[{{Coughlin} {et~al.}(2018{\natexlab{a}}){Coughlin}, {Quataert},
  {Fern{\'a}ndez}, \& {Kasen}}]{Coughlin18a}
{Coughlin}, E.~R., {Quataert}, E., {Fern{\'a}ndez}, R., \& {Kasen}, D.
  2018{\natexlab{a}}, \mnras, 477, 1225

\bibitem[{{Coughlin} {et~al.}(2018{\natexlab{b}}){Coughlin}, {Quataert}, \&
  {Ro}}]{Coughlin18b}
{Coughlin}, E.~R., {Quataert}, E., \& {Ro}, S. 2018{\natexlab{b}}, \apj, 863,
  158

\bibitem[{{Davies} {et~al.}(2011){Davies}, {Hoare}, {Lumsden}, {Hosokawa},
  {Oudmaijer}, {Urquhart}, {Mottram}, \& {Stead}}]{Davies11}
{Davies}, B., {Hoare}, M.~G., {Lumsden}, S.~L., {Hosokawa}, T., {Oudmaijer},
  R.~D., {Urquhart}, J.~S., {Mottram}, J.~C., \& {Stead}, J. 2011, \mnras, 416,
  972

\bibitem[{{Dwarkadas}(2007)}]{Dwarkadas07}
{Dwarkadas}, V.~V. 2007, \apj, 667, 226

\bibitem[{{Ertl} {et~al.}(2016){Ertl}, {Janka}, {Woosley}, {Sukhbold}, \&
  {Ugliano}}]{Ertl16}
{Ertl}, T., {Janka}, H.~T., {Woosley}, S.~E., {Sukhbold}, T., \& {Ugliano}, M.
  2016, \apj, 818, 124

\bibitem[{{Fern{\'a}ndez} {et~al.}(2018){Fern{\'a}ndez}, {Quataert},
  {Kashiyama}, \& {Coughlin}}]{Fernandez18}
{Fern{\'a}ndez}, R., {Quataert}, E., {Kashiyama}, K., \& {Coughlin}, E.~R.
  2018, \mnras, 476, 2366

\bibitem[{{Gerke} {et~al.}(2015){Gerke}, {Kochanek}, \& {Stanek}}]{Gerke15}
{Gerke}, J.~R., {Kochanek}, C.~S., \& {Stanek}, K.~Z. 2015, \mnras, 450, 3289

\bibitem[{{Harris} \& {Zaritsky}(2009)}]{Harris09}
{Harris}, J., \& {Zaritsky}, D. 2009, \aj, 138, 1243

\bibitem[{{Ivanov} \& {Fern{\'a}ndez}(2021)}]{Ivanov21}
{Ivanov}, M., \& {Fern{\'a}ndez}, R. 2021, arXiv e-prints, arXiv:2101.02712

\bibitem[{{Kashiyama} {et~al.}(2018){Kashiyama}, {Hotokezaka}, \&
  {Murase}}]{Kashiyama18}
{Kashiyama}, K., {Hotokezaka}, K., \& {Murase}, K. 2018, \mnras, 478, 2281

\bibitem[{{Kashiyama} \& {Quataert}(2015)}]{Kashiyama15}
{Kashiyama}, K., \& {Quataert}, E. 2015, \mnras, 451, 2656

\bibitem[{{Kleiser} {et~al.}(2011){Kleiser}, {Poznanski}, {Kasen}, {Young},
  {Chornock}, {Filippenko}, {Challis}, {Ganeshalingam}, {Kirshner}, {Li},
  {Matheson}, {Nugent}, \& {Silverman}}]{Kleiser12}
{Kleiser}, I. K.~W., et~al. 2011, \mnras, 415, 372

\bibitem[{{Kochanek}(2014)}]{Kochanek14}
{Kochanek}, C.~S. 2014, \apj, 785, 28

\bibitem[{{Kochanek} {et~al.}(2008){Kochanek}, {Beacom}, {Kistler}, {Prieto},
  {Stanek}, {Thompson}, \& {Y{\"u}ksel}}]{Kochanek08}
{Kochanek}, C.~S., {Beacom}, J.~F., {Kistler}, M.~D., {Prieto}, J.~L.,
  {Stanek}, K.~Z., {Thompson}, T.~A., \& {Y{\"u}ksel}, H. 2008, \apj, 684, 1336

\bibitem[{{Krti{\v{c}}ka} {et~al.}(2021){Krti{\v{c}}ka}, {Kub{\'a}t}, \&
  {Krti{\v{c}}kov{\'a}}}]{Krticka21}
{Krti{\v{c}}ka}, J., {Kub{\'a}t}, J., \& {Krti{\v{c}}kov{\'a}}, I. 2021, \aap,
  647, A28

\bibitem[{{Licquia} \& {Newman}(2015)}]{Licquia15}
{Licquia}, T.~C., \& {Newman}, J.~A. 2015, \apj, 806, 96

\bibitem[{{Lovegrove} \& {Woosley}(2013)}]{Lovegrove13}
{Lovegrove}, E., \& {Woosley}, S.~E. 2013, \apj, 769, 109

\bibitem[{{Maggi} {et~al.}(2016){Maggi}, {Haberl}, {Kavanagh}, {Sasaki},
  {Bozzetto}, {Filipovi{\'c}}, {Vasilopoulos}, {Pietsch}, {Points}, {Chu},
  {Dickel}, {Ehle}, {Williams}, \& {Greiner}}]{Maggi16}
{Maggi}, P., et~al. 2016, \aap, 585, A162

\bibitem[{{Masai}(1984)}]{Masai84}
{Masai}, K. 1984, \apss, 98, 367

\bibitem[{{Masai}(1994)}]{Masai94}
---. 1994, \apj, 437, 770

\bibitem[{{Merloni} {et~al.}(2012){Merloni}, {Predehl}, {Becker},
  {B{\"o}hringer}, {Boller}, {Brunner}, {Brusa}, {Dennerl}, {Freyberg},
  {Friedrich}, {Georgakakis}, {Haberl}, {Hasinger}, {Meidinger}, {Mohr},
  {Nandra}, {Rau}, {Reiprich}, {Robrade}, {Salvato}, {Santangelo}, {Sasaki},
  {Schwope}, {Wilms}, \& {German eROSITA Consortium}}]{Merloni12}
{Merloni}, A., et~al. 2012, arXiv e-prints, arXiv:1209.3114

\bibitem[{{Nadezhin}(1980)}]{Nadyozhin80}
{Nadezhin}, D.~K. 1980, \apss, 69, 115

\bibitem[{{O'Connor} \& {Ott}(2011)}]{OConnor11}
{O'Connor}, E., \& {Ott}, C.~D. 2011, \apj, 730, 70

\bibitem[{{O'Connor} \& {Ott}(2013)}]{OConnor13}
---. 2013, \apj, 762, 126

\bibitem[{{Pastorello} {et~al.}(2012){Pastorello}, {Pumo}, {Navasardyan},
  {Zampieri}, {Turatto}, {Sollerman}, {Taddia}, {Kankare}, {Mattila},
  {Nicolas}, {Prosperi}, {San Segundo Delgado}, {Taubenberger}, {Boles},
  {Bachini}, {Benetti}, {Bufano}, {Cappellaro}, {Cason}, {Cetrulo}, {Ergon},
  {Germany}, {Harutyunyan}, {Howerton}, {Hurst}, {Patat}, {Stritzinger},
  {Strolger}, \& {Wells}}]{Pastorello12}
{Pastorello}, A., et~al. 2012, \aap, 537, A141

\bibitem[{{Raithel} {et~al.}(2018){Raithel}, {Sukhbold}, \&
  {{\"O}zel}}]{Raithel18}
{Raithel}, C.~A., {Sukhbold}, T., \& {{\"O}zel}, F. 2018, \apj, 856, 35

\bibitem[{{Remillard} \& {McClintock}(2006)}]{Remillard06}
{Remillard}, R.~A., \& {McClintock}, J.~E. 2006, \araa, 44, 49

\bibitem[{{Robitaille} \& {Whitney}(2010)}]{Robitaille10}
{Robitaille}, T.~P., \& {Whitney}, B.~A. 2010, \apjl, 710, L11

\bibitem[{{Schure} {et~al.}(2009){Schure}, {Kosenko}, {Kaastra}, {Keppens}, \&
  {Vink}}]{Schure09}
{Schure}, K.~M., {Kosenko}, D., {Kaastra}, J.~S., {Keppens}, R., \& {Vink}, J.
  2009, \aap, 508, 751

\bibitem[{{Smartt} {et~al.}(2009){Smartt}, {Eldridge}, {Crockett}, \&
  {Maund}}]{Smartt09}
{Smartt}, S.~J., {Eldridge}, J.~J., {Crockett}, R.~M., \& {Maund}, J.~R. 2009,
  \mnras, 395, 1409

\bibitem[{{Sukhbold} {et~al.}(2016){Sukhbold}, {Ertl}, {Woosley}, {Brown}, \&
  {Janka}}]{Sukhbold16}
{Sukhbold}, T., {Ertl}, T., {Woosley}, S.~E., {Brown}, J.~M., \& {Janka}, H.~T.
  2016, \apj, 821, 38

\bibitem[{{Tsuna} {et~al.}(2020){Tsuna}, {Ishii}, {Kuriyama}, {Kashiyama}, \&
  {Shigeyama}}]{Tsuna20}
{Tsuna}, D., {Ishii}, A., {Kuriyama}, N., {Kashiyama}, K., \& {Shigeyama}, T.
  2020, \apjl, 897, L44

\bibitem[{{Tsuna} {et~al.}(2018){Tsuna}, {Kawanaka}, \& {Totani}}]{Tsuna18}
{Tsuna}, D., {Kawanaka}, N., \& {Totani}, T. 2018, \mnras, 477, 791

\bibitem[{{Weisskopf} {et~al.}(2002){Weisskopf}, {Brinkman}, {Canizares},
  {Garmire}, {Murray}, \& {Van Speybroeck}}]{Weisskopf02}
{Weisskopf}, M.~C., {Brinkman}, B., {Canizares}, C., {Garmire}, G., {Murray},
  S., \& {Van Speybroeck}, L.~P. 2002, \pasp, 114, 1

\end{thebibliography}

\end{document}